\documentclass[12pt]{article}
\pdfoutput=1
\usepackage[utf8]{inputenc}
\usepackage[T1]{fontenc}
\usepackage{float}
\usepackage{authblk}
\usepackage[normalem]{ulem}
\usepackage{xcolor}
\usepackage{amsmath}
\usepackage{amsthm}
\usepackage{adjustbox}
\usepackage{graphicx}
\usepackage{mathtools}
\usepackage{fullpage}
\usepackage{subcaption}
\usepackage[style=numeric,sorting=none,maxnames=99]{biblatex}
\usepackage[colorlinks = true,
            linkcolor = teal,
            urlcolor  = teal,
            citecolor = teal]{hyperref}

\title{netrd: A library for network reconstruction\\and graph distances}

\author[1]{Stefan McCabe\thanks{mccabe.s@northeastern.edu; klein.br@northeastern.edu}}
\author[1]{Leo Torres}
\author[1]{Timothy LaRock}
\author[1]{Syed Arefinul Haque}
\author[1]{Chia-Hung Yang}
\author[1]{Harrison Hartle}
\author[1,2]{Brennan Klein$^*$}

\affil[1]{Network Science Institute, Northeastern University, Boston, USA}
\affil[2]{Laboratory for the Modeling of Biological and Socio-Technical Systems,\protect\\Northeastern University, Boston, USA}

\date{October 29, 2020}

\begin{document}
\maketitle
\pagenumbering{arabic}

\section{\label{sec:introduction}Summary and package description}

Complex systems throughout nature and society are often best represented as \textit{networks}. Over the last two decades, alongside the increased availability of large network datasets, we have witnessed the rapid rise of network science \cite{Amaral2004, Vespignani2008, Newman2010, Barabasi2016}. This field is built around the idea that an increased understanding of the complex structural properties of a variety systems will allow us to better observe, predict, and even control the behavior of these systems.

However, for many systems, the data we have access to is not a direct description of the underlying network. More and more, we see the drive to study networks that have been inferred or reconstructed from non-network data---in particular, using \textit{time series} data from the nodes in a system to infer likely connections between them \cite{Brugere2018, Runge2018}. Selecting the most appropriate technique for this task is a challenging problem in network science. Different reconstruction techniques usually have different assumptions, and their performance varies from system to system in the real world. One way around this problem could be to use several different reconstruction techniques and compare the resulting networks. However, network comparison is also not an easy problem, as it is not obvious how best to quantify the differences between two networks, in part because of the diversity of tools for doing so.

The \texttt{netrd} Python package seeks to address these two parallel problems in network science by providing, to our knowledge, the most extensive collection of both network reconstruction techniques and network comparison techniques (often referred to as \textit{graph distances}) in a single library (\url{https://github.com/netsiphd/netrd/}). In this article, we detail the two main functionalities of the \texttt{netrd} package. Along the way, we describe some of its other useful features. This package builds on commonly used Python packages (e.g. \texttt{networkx} \cite{SciPyProceedings11}, \texttt{numpy} \cite{Harris2020}, \texttt{scipy} \cite{SciPy2020}) and is already a widely used resource for network scientists and other multidisciplinary researchers. With ongoing open-source development, we see this as a tool that will continue to be used by all sorts of researchers to come.

\subsection{\label{sec:reconstructions}Network reconstruction from time series data}

\begin{figure}[t!]
    \centering
    \includegraphics[width=1.0\columnwidth]{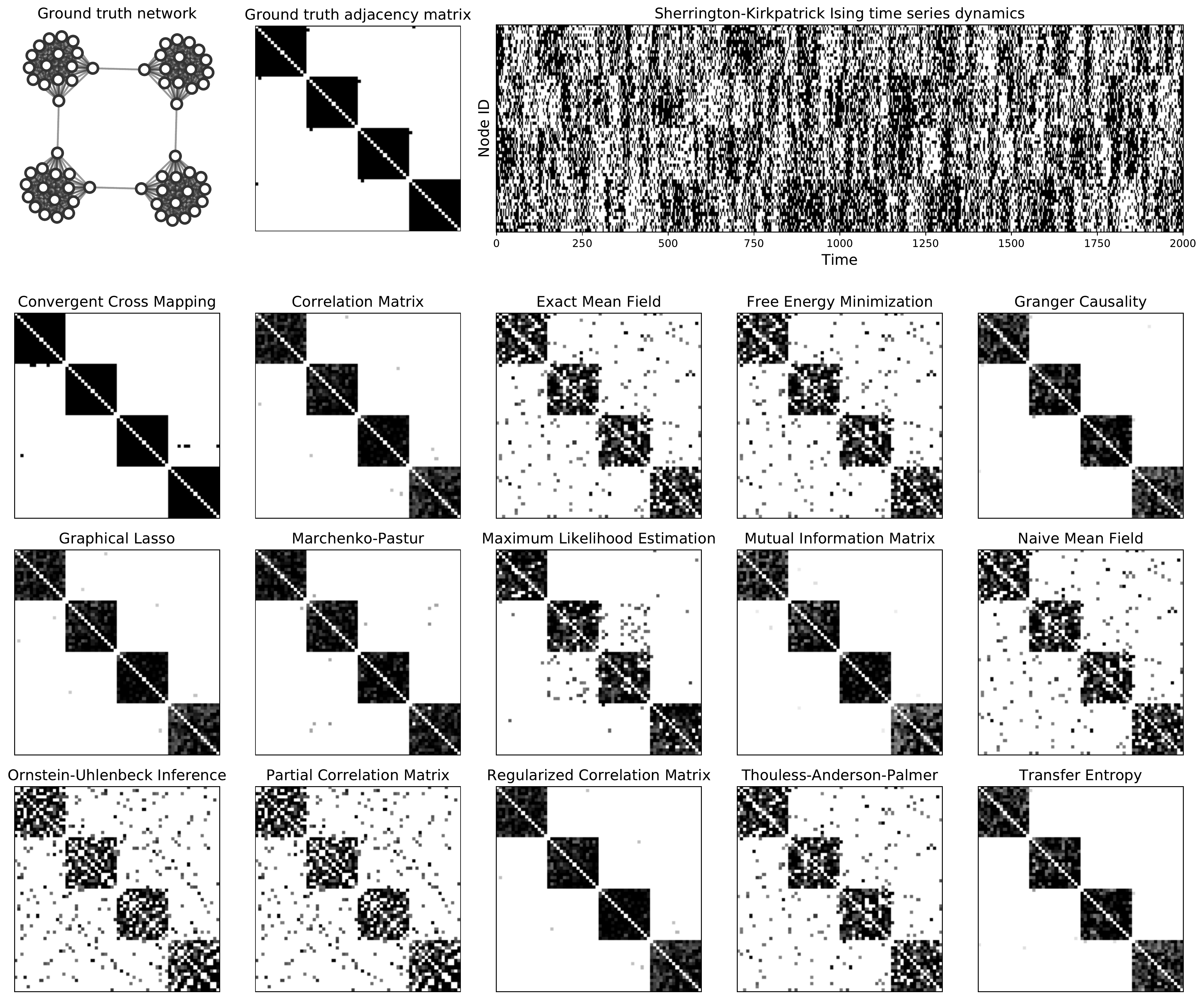}
    \caption{\textbf{Example of the network reconstruction pipeline.} (Top row) A sample network, its adjacency matrix, and an example time series, $TS$, of node-level activity simulated on the network. (Bottom rows) The outputs of 15 different network reconstruction algorithms, each using $TS$ to create a new adjacency matrix that captures key structural properties of the original network.}
    \label{fig:ground}
\end{figure}

Given time series data, $TS$, of the behavior of $N$ nodes / components / sensors of a system over the course of $L$ timesteps, and given the assumption that the behavior of every node, $v_i$, may have been influenced by the past behavior of other nodes, $v_j$, there are dozens of techniques that can be used to infer which connections, $e_{ij}$, are likely to exist between the nodes. That is, we can use one of many \textit{network reconstruction} techniques to create a network representation, $G_r$, that attempts to best capture the relationships between the time series of every node in $TS$. \texttt{netrd} is a Python package that lets users perform this network reconstruction task using 17 different techniques, meaning that many different networks can be created from a single time series dataset. For example, in Figure \ref{fig:ground} we show the outputs of 15 different reconstruction techniques applied to time series data generated from an example network \cite{Sugihara2012, Mishchenko2011, Hoang2019, Sheikhattar2018, Friedman2008, Edelman2005, Zeng2013, Donges2009, Barucca2014, Ledoit2003, Stetter2012, Peixoto2019}.

\subsubsection{\label{sec:dynamics}Simulated network dynamics}

Practitioners often apply these network reconstruction algorithms to real time series data. For example, in neuroscience, researchers often try to reconstruct functional networks from time series readouts of neural activity \cite{Mishchenko2011}. In economics, researchers can infer networks of influence between companies based on time series of changes in companies' stock prices \cite{Squartini2018}. At the same time, it is often quite helpful having the freedom to \textit{simulate} arbitrary time series dynamics on randomly generated networks. This provides a controlled setting to assess the performance of network reconstruction algorithms. For this reason, the \texttt{netrd} package also includes a number of different techniques for simulating dynamics on networks.

\subsection{\label{sec:distances}Comparing networks using graph distances}

\begin{figure}[t!]
    \centering
    \includegraphics[width=1.0\columnwidth]{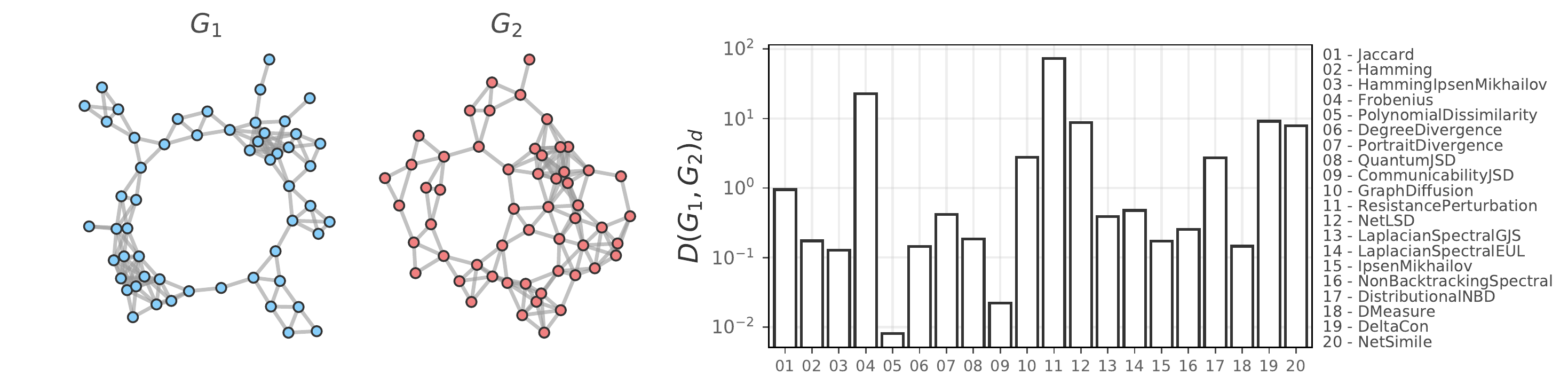}
    \caption{\textbf{Example of the graph distance measures in \texttt{netrd}.} Here, we measure the graph distance between two networks using 20 different distance measures from \texttt{netrd}.}
    \label{fig:dists}
\end{figure}

A common goal when studying networks is to describe and quantify how different two networks are. This is a challenging problem, as there are countless axes upon which two networks can differ; as such, a number of \textit{graph distance} measures have emerged over the years attempting to address this problem. As is the case for many hard problems in network science, it can be difficult to know which (of many) measures are suited for a given setting. In \texttt{netrd}, we consolidate over 20 different graph distance measures into a single package \cite{Jaccard1901, Hamming1950, Jurman2015, Golub2013, Donnat2018, Carpi2011, Bagrow2019, DeDomenico2016, Chen2018, Hammond2013, Monnig2018, Tsitsulin2018, Jurman2011, Ipsen2002, Torres2019, Mellor2019, Schieber2017, Koutra2016, Berlingerio2012}. Figure \ref{fig:dists} shows an example of just how different these measures can be when comparing two networks, $G_1$ and $G_2$. This submodule in \texttt{netrd} has already been used in recent work with a novel characterization of the graph distance literature \cite{Hartle2020}.

\section{\label{sec:other}Related software packages}

In the network reconstruction literature, there are often software repositories that detail a single technique or a few related ones. For example Lizier (2014) implemented a Java package (portable to Python, octave, R, Julia, Clojure, MATLAB) that uses information-theoretic approaches for inferring network structure from time-series data \cite{Lizier2014}; Runge et al. (2019) created a Python package that combines linear or nonlinear conditional independence tests with a causal discovery algorithm to reconstruct causal networks from large-scale time series datasets \cite{Runge2019}. These are two examples of powerful and widely used packages though neither includes as wide-ranging techniques as \texttt{netrd} (nor were they explicitly designed to). In the graph distance literature, the same trend is broadly true: many one-off software repositories exist for specific measures. However, there are some packages that do include multiple graph distances; for example, Wills (2017) created a \texttt{NetComp} package that includes several variants of a few distance measures included here \cite{Wills2017}.

\paragraph{Acknowledgements} 

The authors thank Kathryn Coronges, Mark Giannini, and Alessandro Vespignani for contributing to the coordination of the 2019 Network Science Institute ``Collabathon'', where much of the development of this package began. The authors acknowledge the support of ten other contributors to this package: Guillaume St-Onge, Andrew Mellor, Charles Murphy, David Saffo, Carolina Mattsson, Ryan Gallagher, Matteo Chinazzi, Jessica Davis, Alexander J. Gates, and Anton Tsitulin. \textbf{Funding:} This research was supported by the Network Science Institute at Northeastern University. B.K. is supported by the National Defense Science \& Engineering Graduate Fellowship (NDSEG) Program.

\printbibliography[title={References}]

\end{document}